\documentclass[lettersize,journal]{IEEEtran}
\usepackage{amsmath,amsfonts}
\usepackage{algorithm}
\usepackage{algpseudocode}
\usepackage{array}
\usepackage[caption=false,font=normalsize,labelfont=sf,textfont=sf]{subfig}
\usepackage{textcomp}
\usepackage{stfloats}
\usepackage{url}
\usepackage{verbatim}
\usepackage{graphicx}
\usepackage{cite}
\usepackage{multirow}
\usepackage[table,xcdraw]{xcolor}

\begin{document}

\title{MODRIC: A Cost Effective MODular Data Center Network Architecture with Rich InterConnections}
\author{
    Nabajyoti Medhi\IEEEauthorrefmark{1}, 
    Kumarjit Ray\IEEEauthorrefmark{2}, 
    Rajdeep Ghosh\IEEEauthorrefmark{3}, 
    Dilip Kumar Saikia\IEEEauthorrefmark{4} \\
    
    \IEEEauthorblockA{\IEEEauthorrefmark{1}\IEEEauthorrefmark{4}Department of Computer Science \& Engineering, Tezpur University, India} 
    
    \IEEEauthorblockA{\IEEEauthorrefmark{2}\IEEEauthorrefmark{3}Department of Computer Science \& Engineering, Indian Institute of Technology Kharagpur, India \\
    Email: \IEEEauthorrefmark{1}nmedhi@tezu.ernet.in, \IEEEauthorrefmark{2}kumarjit@kgpian.iitkgp.ac.in, \IEEEauthorrefmark{3}rajdeep\_csb18@agnee.tezu.ernet.in, \IEEEauthorrefmark{4}dks@tezu.ernet.in
    } 
    
}



\maketitle

\begin{abstract}
Shipping container based modular architectures provide design flexibility in data centers with building blocks to expand the network as and when needed. In this paper, high capacity Modular Data Center (MDC) network architecture with Rich Inter Connections named MODRIC is proposed. MODRIC is a cost-effective switch-centric network design which allows building a flexible MDC network with commodity switches. It uses an inter-container connectivity similar to the structure of generalized hypercube in order to provide high inter-container bandwidth. Further, a hybrid Clos topology is used to build the container network. MODRIC is highly suitable for cost effectively building mega data centers requiring high throughput capacity and resilience against failures. This paper presents the proposed architecture, discusses its relevant properties, and proposes suitable addressing, routing and network construction schemes. The paper also presents comparative studies on its cost and performance with existing network topologies.
\end{abstract}


\section{Introduction}
\IEEEPARstart{D}{ata} center networks are generally used to provide large scale computing resources in the form of servers and storage available on the internet \cite{9925714}. These generally use distributed file systems \cite{apache_software_foundation_2019} and distributed execution engines \cite{apache_software_foundation_2019, dean2004mapreduce} to enable very large scale data processing. Large data centers are deployed by the companies which provide cloud services \cite{varia_mathew_2014} or other popular online web services. As the usage grows for a data center, the requirement in the number of servers increases \cite{10535462}. Addition of servers requires expansion of the network with more cables and switches for connectivity. While doing so, the topological properties of the network should remain intact \cite{jia2023sretor}. Else there will be degradation in performance of the network. For time critical applications, even a small delay in response can affect the overall performance of the application very badly. Therefore addition of devices to the network needs to be done carefully. To enable more structured expansion to the network many companies are shifting towards building shipping container based modular data centers (MDC) \cite{sun} which are portable and easier to manage. In MDC network, a shipping container \cite{sun} may contain hundreds of servers placed on multiple racks. The racks with switches and servers are placed inside the container in such a manner that they can be easily managed in terms of cabling, powering and monitoring. Such containers can be added to the MDC to incrementally upgrade the network. However challenges exist in building MDCs due to the need for high inter-container bandwidth, cost of interconnection devices and possible complexity in cabling.  

In designing an MDC, there are several goals that must be achieved: 

\textbf{High scalability:} MDC network architecture needs to provide high capacity to expand the network in terms of number of servers to enable enhancement in the computational power with negligible degradation in performance. There should be flexibility in the granularity for expansion.

\textbf{Low cabling complexity:} Cabling complexity affects the scalability of the DCN with the increase in network size. Rich cabling is often necessary in order to provide multipath and resilience. Cabling within a network rack is easy to maintain whereas inter-rack cabling should be done carefully when a large number of racks are inter-connected. In an MDC, inter-container cabling brings challenge when the number of containers increases. This is due to the need of arranging long inter-container cables. The MDC topology should allow the design of a structured inter-container cabling such that long cables can be grouped into bundles for easier management.

\textbf{High inter-container throughput:} Data center network (DCN) traffic goes high for many cloud-scale high bandwidth applications. A major proportion of the overall traffic, in such applications, consists of east-west traffic i.e., traffic between VMs/ servers within the data center. As a result, inter-container host-to-host traffic accounts for a large portion of the MDC network traffic. Several distributed execution engines like MapReduce \cite{dean2004mapreduce} even generates all-to-all host traffic where each host sends traffic to all the other hosts. This calls for high capacity interconnections within the DCN. Further, use of high capacity optical fiber links for inter-container connections overcomes the propagation distance limitation of copper cables. 

\textbf{Low latency:} In order to minimize the latency, a network topology is desired to have small network diameter \cite{kotsis1992interconnection}. Provisions also need to be made for parallel paths through rich interconnections to avoid delays due to congestions.

\textbf{Fault-tolerance:} Existence of multiple parallel paths between source-destination pairs of hosts supported by suitable fault-tolerant routing strategy provides good fault tolerance in a network. Richer the inter-container network connectivity, higher is the likelihood of fault tolerance of the MDC.

\textbf{Decoupled inter-container and intra-container network:} The decoupling allows- heterogeneity in the containers, better manageability, technology independent updates in the containers, change in inter-container network design without affecting intra-container network and use of separate sets of protocols for the two networks.

\textbf{Cost effectiveness:} The DCN cost depends mainly on costs of switching devices, servers and cabling. Because of the large size, to be cost-effective the DCN needs to be amenable to use commodity devices in order to minimize the overall cost. 
In this paper, we propose MODRIC, a flexible, cost-effective and high capacity switch-centric structured MODular data center design with Rich InterConnections to achieve the design goals stated above. The paper presents the design of the proposed architecture, discusses its relevant properties, provides suitable addressing, routing and network construction schemes. The paper also presents comparative study on its cost of implementation. A performance comparison with existing modular DCN topologies is also presented which has been performed in a real software switch based virtual network platform.

The organization of the paper is as follows. Section 2 discusses the related works; section 3 discusses the proposed architecture of MODRIC and section 4 deals with some topological and structural properties of MODRIC interconnection network. Section 5 discusses the design issues for construction and expansion of the network, section 6 illustrates the networking in MODRIC, section 7 discusses the design and cost analysis of MODRIC network, section 8 presents the performance evaluation and simulation results. Finally we conclude the paper in section 9.

\section{RELATED WORKS}
In recent years there have been significant amount of work done on data centre networks (DCN) architecture. In this section, we present some significant works done in the field of modular data center architecture.

The Fat-tree \cite{88} is one of the most popular designs for data center networks. It is a type of Clos topology \cite{https://doi.org/10.1002/j.1538-7305.1953.tb01433.x} which provides low oversubscription ratio at all the switches in the data center. Clos topologies provide high bisection bandwidth with non-blocking connections. 

Jellyfish \cite{180604} uses random regular graph for incremental expansion of the network. Jellyfish supports higher network capacity with larger number of servers. But cabling in Jellyfish is unstructured and complex which makes it unsuitable for modular data centers.  

BCube \cite{guo2009bcube} is designed to build shipping-container based data centers. It uses a server centric approach using a recursive structure. It also supports graceful performance degradation as a host or switch failure occurs. BCube provides multiple parallel shortest paths between any pair of hosts. In BCube, switches at different levels communicate via servers and this overburdens the servers with extra relaying responsibility. 

MDCube \cite{10.1145/1658939.1658943} gives a structured modular data center design with BCube containers. The inter-container connectivity in MDCube is based on generalized hypercube \cite{1676437} architecture. Each container has a limited number of high speed links to other containers depending upon the number of high speed interfaces in a conventional COTS (Commodity off the Shelf) switch. Due to the use of only low cost COTS switches for inter-container connections MDCube provides a low cost solution for building MDCs. 

uFix \cite{6089059} proposes a mechanism to build a mega data center with heterogeneous containers. uFix focuses on designing the interconnection in such a way that the servers or switches need not be upgraded with the network expansion. uFix inter-container connection and routing are decoupled from the intra-container network. However, uFix emphasizes on server-centric inter-connection among containers since servers provide programming flexibility and leads to less operational cost. uFix deals with a server-centric inter-container connectivity with an irregular cabling structure.

Server-centric designs face capacity limitations in networks that need high routing capacity due to voluminous server to server traffic. The servers therefore need high capacity NICs for faster forwarding but face the limitations of available speeds. Additional CPU cores are required in servers \cite{10.1145/1921168.1921189} to overcome the negative impact of routing load on the applications. Upgrading servers with more NICs is time consuming. Switches with thin ports can perform much better than relay servers in the routing of packets \cite{10.1145/1250662.1250679}. Hence switch centric networks have clear advantage.

\section{MODRIC DESIGN}
The MODRIC MDC network architecture is designed at two levels- the inter-container network and the intra-container network (or, container network). It uses inter-container connectivity similar to a generalized hypercube \cite{1676437} in a two dimensional $m \times n \quad (m \geq 1, \, n \geq 1)$ grid formation of the containers. The intra-container network on the other hand has a 2 layered Clos like topological structure. As per the design objectives the two networks are decoupled. The design can exploit the merchant silicon \cite{5238680} based commodity switches to achieve the set goals. 

The links used in the network are of two speeds- one with very high capacity C denoted as S1 (say, 10GbE) and the other with lower capacity c denoted as S2 (say, 1GbE). The S2 links will be copper links while S1 links may be copper or optical fiber depending upon the length.

In the MODRIC design, a traffic constraint is followed so that 1:1 oversubscription can be achieved at the switches interfacing the containers as well at those within the container network.

\subsection{MODRIC Container Network}
   
    \begin{figure}[!t]
        \begin{center}
            \includegraphics[width=3.2in]{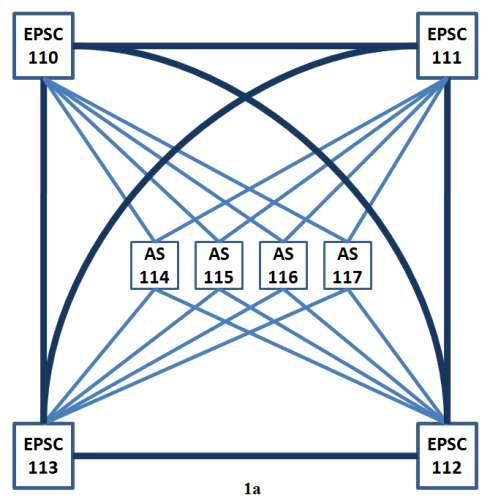}
            \includegraphics[width=3.2in]{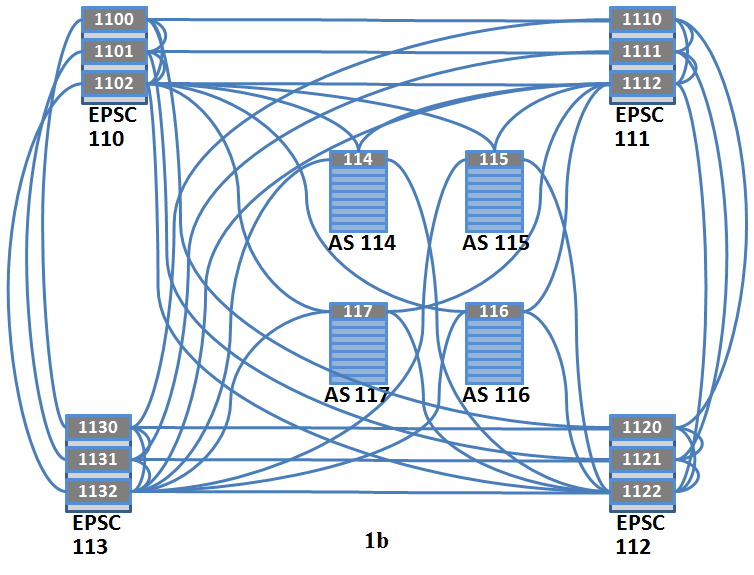}
            \caption{Top down canonical view of a MODRIC container (cid = 11) with four ASes (114, 115, 116, 117). a. Graphical view showing the EPSs and ASes and the interconnections. Inter-EPSC S1 links between a pair of EPSCs are shown as a single thick dark link. ASes are connected to EPSCs through S1 links.      b. View of the container where ASes are located on top of racks containing the servers connected to each AS (blue shelves, S2 links are not shown) and each EPSC containing three EPSes connected to each other. All the connections shown are with S1 links.}
            \label{fig_1}
        \end{center}
    \end{figure}

A MODRIC container network has two parts- a Container Boundary Network (CBN) and a Container Internal Network (CIN).

The CBN in each container (Figure \ref{fig_1}) in MODRIC has a set of four end-point switch clusters (EPSC) each having one or more end-point switches (EPS) depending on the number of port required. The four EPSCs are ranked $r = 0, 1, 2, 3$ clock-wise and assigned EPSC ids with ranks prefixed by the Container id (cid) as shown in Figure \ref{fig_1}.

The EPSes in an EPSC are connected using S1 copper links forming a full mesh within the EPSC in order to provide one-hop reachability among them for the EPSC to effectively act as a single modular switch unit. Initially, each EPSC may have only one EPS. With the increasing size of a container and/or the MDC, number of EPSes in the EPSCs may be increased as the requirement of number of ports increases in the EPSC. This makes the overall network expansion easier and thus makes the network highly scalable.

The EPSes inside an EPSC can be further ranked with a fourth digit l where l represents EPS layer in the EPSC e.g., 1120 represents the EPS in layer 0 (top-most layer with l = 0) of EPSC 112. The EPSes in the same layer of the four EPSCs also form full mesh using S1 copper links to provide one-hop reachability among the same layered EPSes in the four EPSCs. 

The Container Internal Network (CIN) comprises a tree with a set of x ToR Access Switches (AS) and a set of y Host Servers (H) connected to each of the ASes with S2 links. Each of the ASes is connected to one EPS in each of the four EPSCs inside the container via S1 links. The ASes are ranked $r = 4, 5, 6, \dots$  with the first 4 ranks assigned to the EPSCs in the container. AS ids comprise the individual AS ranks prefixed by the Container Id (cid) as shown in Figure \ref{fig_1}.

The traffic constraint followed in the CIN can be stated as - \textit{the aggregate capacity of the intra-container links from an AS to the hosts does not exceed the aggregate capacity of the links from the AS to the four EPSCs. This constraint results in the relation: $y \leq \frac{4C}{c}$.}

Example: For a container network with 40 host servers per AS, a 42U rack can be used for placing the AS and the servers. A commodity TOR1G switch with 4x10GBE uplink ports and 40x1GbE downlink ports can be taken as the AS. The 40 servers connected to the AS with GbE copper cables.

\subsection{MODRIC Inter-container Network}
\begin{figure}[!t]
        \begin{center}
            \includegraphics[width=3.2in]{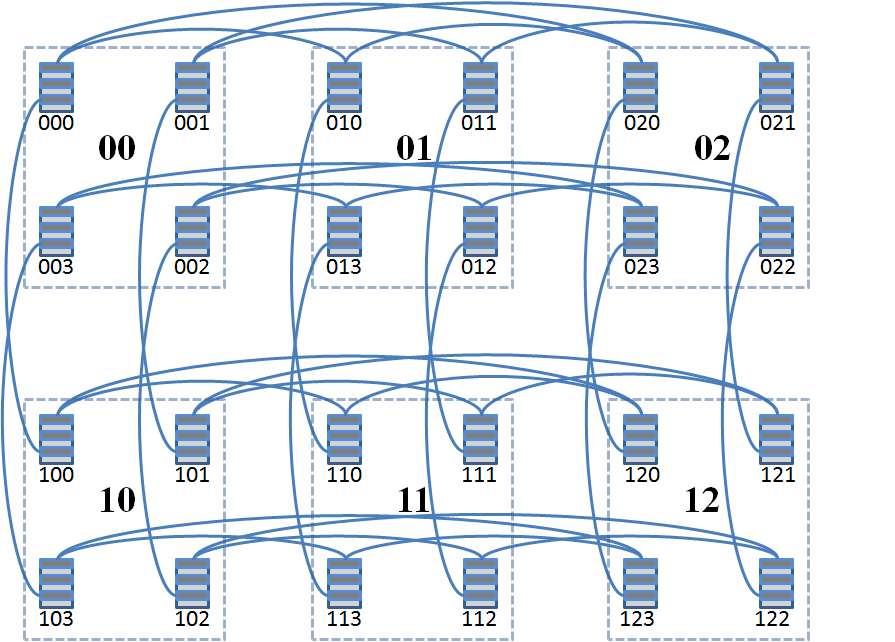}
            \caption{A 2x3 MODRIC network (showing the containers with EPSCs and the inter-container connectivity. The numbers with larger digits are container ids and the ones with smaller digits are EPSC ids)}
            \label{fig_2}
        \end{center}
    \end{figure}

The inter-container network comprises the connectivity between EPSCs in the containers that are arranged in a 2-dimensional \(m \times n\) grid formation [Figure \ref{fig_2}]. An EPSC in one container is connected to the EPSC of the same rank located in each of the containers in the same row and the same column with S1 links. Each EPSC will have \(m - 1\) adjacent EPSCs in the column and \(n - 1\) adjacent EPSCs in the row for a total of \(m+n-2\) adjacent EPSCs in the network.

The traffic constraint for the MODRIC inter-container network can be stated as- the aggregate capacity of all the intra-container links from an EPSC to the access switches does not exceed the aggregate capacity of all the inter-container links from that EPSC. As per this constraint, at least x inter-container links are required from an EPSC. If \(x > m + n - 2\), which may occur in cases of small values of m and n, additional links will be added between the adjacent EPSCs. These may be added along the rows or the columns.

The inter-EPSC cabling in the CBN can be done between any pair of free ports of the adjacent EPSCs.

\section{TOPOLOGICAL AND STRUCTURAL PROPERTIES OF MODRIC}
\begin{figure}[!t]
        \begin{center}
            \includegraphics[width=3.2in]{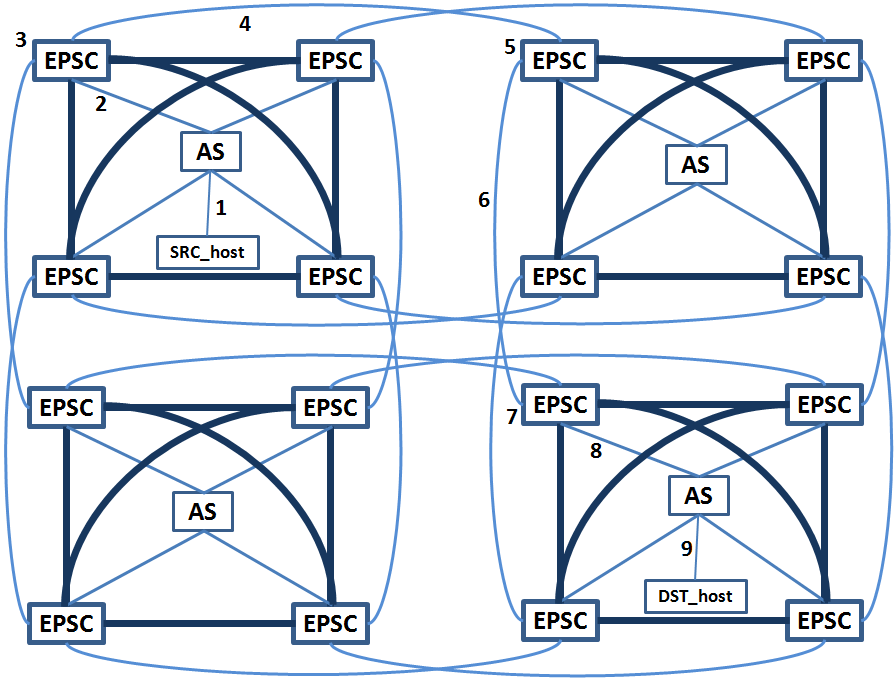}
            \caption{Network diameter of MODRIC shown with 9 points between a pair of source and destination hosts. Points 3, 5, and 7 contribute one hop distance each due to EPSC connectivity.}
            \label{fig_3}
        \end{center}
    \end{figure}

In this section, we discuss three important topological properties of MODRIC that address the design objectives of an MDC network. We also discuss three structural properties that are related to the size of a container and the cabling complexity of MODRIC.

The diameter of a network \cite{kotsis1992interconnection} is defined as the maximum of the shortest distances among all pairs of its server nodes. A smaller diameter means lower latency. Constant diameter avoids the possibility of degradation in performance with the expansion of the network and thus enhances scalability.\\

\textbf{Property 1:} \textit{The diameter of a MODRIC network is 9.} \par

\textbf{Proof:}
An access switch is 1 hop away from each of the EPSCs in the container. An EPSC in a container is 1 hop away from at least one of the EPSCs in any of the containers in the network that are in the same row or column as its own. Further, each EPSC contributes 1 hop distance due to the intra-EPSC mesh connectivity inside. Hence, any two access switches in two different containers that are in the same row or column are within 5 hops from each other.  However, if the two access switches are not in containers that are in the same row or column, there will be a need for two additional hops via a container that shares the row or column with the two containers. Hence, the distance between the two access switches will be 7 in this case, and thereby, the distance between any two hosts connected to these two access switches each will be 9 [Figure \ref{fig_3}].

Therefore, MODRIC has a small constant diameter of 9 for any dimensions m and n of the MDC network. This ensures lower latency in routing in MODRIC network.\\

\textbf{Property 2:} \textit{The number of node-disjoint parallel shortest paths between any two ASes in two different containers is 4.}\par

\textit{The number of node-disjoint parallel shortest paths between any two containers is-}
\textit{\renewcommand{\labelenumi}{\alph{enumi})}
\begin{enumerate}
    \item 4, if the two containers are in the same row or column.
    \item 8, if the two containers are not in the same row or column.
\end{enumerate}} \par

\textbf{Proof:}\par 
\renewcommand{\labelenumi}{\alph{enumi})}
\begin{enumerate}
    \item \textbf{Node-disjoint paths between ASes:} From an AS, there are direct links to the 4 EPSCs within the container. The 4 EPSCs are connected by 4 parallel links to the 4 EPSCs in the next container in the path to the destination AS. These provide 4 node-disjoint paths between any two ASes.

    \item \textbf{Node-disjoint paths between Containers:} There are 4 parallel links between any two containers in the same row or column through the 4 EPSCs to create 4 alternative paths between the two containers, and as these consist of separate single hop links, the paths are node-disjoint. These also lead to each of the ASes in the two containers through 4 separate links so that there are 4 node-disjoint paths between any two ASes in containers in the same row or column. 

    In case the two containers are not in the same row/ column, the shortest path can be taken with a row hop first and a column hop next or the reverse, with 4 alternative links in both directions. It leads to 2x4=8 alternative inter-container node-disjoint parallel paths. 
    
    However, between the EPSCs and an AS, there are only 4 links. Therefore the number of node-disjoint paths between any two ASes in located in two containers remains 4.
    
    Due to the availability of node/edge-disjoint multiple shortest paths, MODRIC guarantees high inter-container throughput and high fault tolerance. \\
\end{enumerate}

\textbf{Property 3:} \textit{The bisection bandwidth of a MODRIC container is \((2x + 4q) \cdot C\), where \(q\) is the number of EPSes per EPSC and \(C\) is the capacity of S1 links.} \par
\textit{The bisection bandwidth of a MODRIC network of dimension \(m \times n\) (where \(1 < m \leq n\)) is \(O(n \cdot m^2 \cdot C)\).} \par

\textbf{Proof: }
The proof is shown in Appendix A.\par 
Bisection bandwidth is the worst-case bandwidth between two equal halves of the network. High bisection bandwidth ensures a high-capacity network. For MODRIC, it maximizes to \(n^3 C\) for \(m = n\). The bisection bandwidth of MODRIC grows approximately with the cubic power of its dimension, making it a highly scalable topology for MDCs.

\textbf{Property 4:} \textit{The maximum number of hosts in a MODRIC container is \((m + n - 2) \cdot \left(\dfrac{4C}{c}\right)\), and that in an \(m \times n\) MODRIC network is \(m \cdot n \cdot (m + n - 2) \cdot \left(\dfrac{4C}{c}\right)\), when the restriction \(x \leq (m + n - 2)\) is adhered to.}

\textbf{Proof:}
There are \(x\) ASes in a container and \(y\) hosts connected to each of the ASes. Therefore, the number of hosts in a MODRIC container is \(x \cdot y\). Due to traffic constraints, the upper limits on \(x\) and \(y\) are \((m + n - 2)\) and \(\dfrac{4C}{c}\), respectively. Thus, the upper limits on the number of hosts in a MODRIC container and an \(m \times n\) MODRIC network are \((m + n - 2) \cdot \dfrac{4C}{c}\) and \(m \cdot n \cdot (m + n - 2) \cdot \dfrac{4C}{c}\), respectively.

\textbf{Property 5:} 
\textit{The maximum number of S2 link switch ports required in an \(m \times n\) MODRIC network is \(m \cdot n \cdot (m + n - 2) \cdot \dfrac{4C}{c}\).}  \par
\textit{The maximum number of S1 link switch ports required in an \(m \times n\) MODRIC network is \(4m \cdot n \cdot \left(q^2 + 2q + 3m + 3n - 6\right)\), where \(q\) is the number of EPSes in an EPSC.}

\textbf{Proof:}
The S2 links between the ASes and the hosts are used in MODRIC. Therefore, the number of S2 link switch ports required in the network is equal to the number of hosts in it. Thus, the upper limit on the number of S2 link switch ports in an \(m \times n\) MODRIC network is \(m \cdot n \cdot (m + n - 2) \cdot \dfrac{4C}{c}\).

The S1 links are used within a MODRIC container for the mesh of \(q\) EPSes within each EPSC and the mesh of EPSes in the same layer in the 4 EPSCs in the container. S1 links are also used between each EPSC and \(x\) ASes connected to it, requiring a total of \(8x\) S1 link ports in a container, which includes a total of \(4x\) ports of 4 EPSes and \(4x\) ports of \(x\) ASes. This leads to a maximum of \(4q \cdot \left( (q - 1) + 3 \right) + 8x = 4 \left( q \cdot (q + 2) + 2(m + n + 2) \right)\) S1 link switch ports for the intra-container links in a container.

All the inter-container links in the MODRIC network are S1 links. Each of the 4 EPSCs in a container will have \((m + n - 2)\) S1 ports, leading to a total of \(4(m + n - 2)\) S1 link switch ports in a container.

The total number of S1 link switch ports required in an \(m \times n\) MODRIC network is therefore
\begin{align*}
4m \cdot n \cdot \Big( q \cdot (q + 2) &+ 2(m + n - 2) \\
&+ (m + n - 2) \Big) = \\ &4m \cdot n \cdot \Big( q^2 + 2q + 3m + 3n - 6 \Big).
\end{align*}

\textbf{Property 6:}
\textit{The length of external fiber-optic cables needed to connect an \(m \times n\) MODRIC network in a two-dimensional matrix arrangement of the containers is 
\[
\frac{2}{3} \cdot \left( m \cdot (n^3 - n) \cdot (L + dL) + n \cdot (m^3 - m) \cdot (W + dW) \right),
\]
where \(L\) and \(W\) are the length and width of a container, and \(dL\) and \(dW\) are the distances between two containers in the row direction and column direction, respectively.}

\textbf{Proof:}
\begin{figure}[!t]
        \begin{center}
            \includegraphics[width=2.2in]{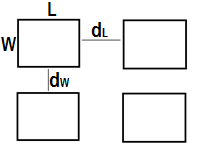}
            \caption{Each container is of length L and width W. Adjacent containers are spaced at dL and dW distance apart along the row and column directions, respectively.}
            \label{fig_4}
        \end{center}
    \end{figure}
The fiber-optic connectivity in MODRIC is limited only to inter-container links, and the inter-container links are only between the EPSCs of the same ranks in containers in the same row and column. For simplicity, let us take the canonical MODRIC inter-container structure shown in Fig. 2 and assume that all the containers in the MDC have similar physical arrangements of the hosts and the switches. Then, the distance between EPSCs on the same rank in adjacent containers in the same row will be \(L + dL\), and that in the case of containers in the same column will be \(W + dW\) as shown in Figure \ref{fig_4}.

Let us consider the inter-container connections for EPSCs in a row. Considering the inter-connections between the EPSCs of a particular rank in the row, the total length of these connections can be expressed as
\begin{align*}
((n - 1) \cdot 1 + (n - 2) \cdot 2 + \dots + 1 \cdot (n - 1)) \cdot (L + dL) &= \\
\frac{(n^3 - n) \cdot (L + dL)}{6}.
\end{align*}

For all 4 EPSCs, the total length will be
\[
\frac{2}{3} \cdot (n^3 - n) \cdot (L + dL).
\]

For the \(m\) rows of containers, the total length of the inter-container cables along the rows shall be
\[
\frac{2}{3} \cdot m \cdot (n^3 - n) \cdot (L + dL).
\]

Similarly, the total length of inter-container cables along the columns shall be
\[
\frac{2}{3} \cdot n \cdot (m^3 - m) \cdot (W + dW).
\]

Thus, the total inter-container cable length is
\[
\frac{2}{3} \cdot \left( m \cdot (n^3 - n) \cdot (L + dL) + n \cdot (m^3 - m) \cdot (W + dW) \right).
\]


\section{DESIGN ISSUES FOR CONSTRUCTION}
In this section, we discuss the design issues of expansion and cabling of a MODRIC network. 

\subsection{Network expansion}

MODRIC provides the feature of modular expansion, which allows very large-scale expansion of the DCN. As per the design objectives, the container network is decoupled from the inter-container network. This makes expansion of the network possible by expanding one or more containers or by adding containers to the network. 

The container expansion is possible by adding host servers to the ASes or by adding ASes to the container. The number of servers that can be added to an AS is restricted to \(y \leq \frac{4C}{c}\) or the number of ports available in the AS, whichever is lower. The number of ASes (\(x\)) in the container is restricted to the number of inter-container links from an EPSC or by the number of ports available in the EPSC. These restrictions can be overcome by adding additional inter-container links to the EPSCs and adding EPSes to EPSCs, respectively. 

In a container, each EPSC may initially consist of only one EPS. The number of EPSes is increased evenly in each of the four EPSCs of the container when there are no more free ports remaining for the expansion. 

Expansion of the inter-container network can also be done by adding rows or columns of containers to the network as required. Additions can be made to an existing \(m \times n\) MDC network for another row with \(n\) containers or with another column with \(m\) containers to form either a \((m + 1) \times n\) or an \(m \times (n + 1)\) network. This naturally results in a larger increment in the size of the network.

\subsection{Cabling complexity and packaging}
\begin{figure}[!t]
    \begin{center}
        \subfloat[Packaging cables between EPSCs and ASes. Four EPSCs are located in one place to form a single EPSC cluster (EPSCC)]{
            \includegraphics[width=3.2in]{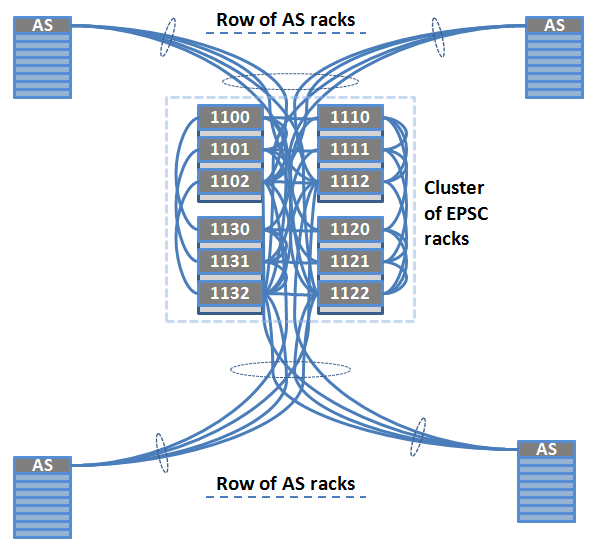}
            \label{fig_5a}
        } \\
        \subfloat[EPSC cluster (EPSCC) is placed in a single large EPSCC rack]{
            \includegraphics[width=3.2in]{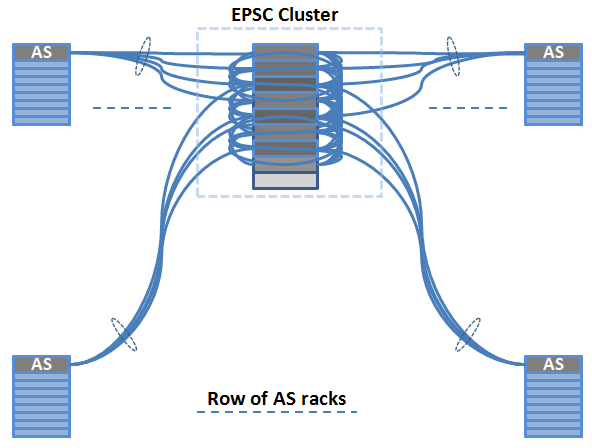}
            \label{fig_5b}
        }
        \caption{Illustration of the MODRIC network setup}
        \captionsetup[subfigure]{font=normalsize} 
        \label{fig_5}
    \end{center}
\end{figure}

Cabling complexity is a major issue in the construction of a DCN. Usually, multiple cables between any two devices, racks, or containers are bundled into a single bunch for ease of management. The devices are also grouped in the container network to minimize the cabling overhead. 

In a container, EPSC racks can be placed together as a single unit to form an EPSC cluster (EPSCC). The EPSCC will send out two bundles of cables to connect two rows of ASes within the container (Figure \ref{fig_5a}). If the number of EPSes per EPSC is small, then all four EPSCs can be placed in a single rack (EPSCC rack) as shown in Figure \ref{fig_5b}.   

\begin{figure}[!t]
    \begin{center}
        \subfloat[Packaging of inter-container cables between two containers into two bundles]{
            \includegraphics[width=3.2in]{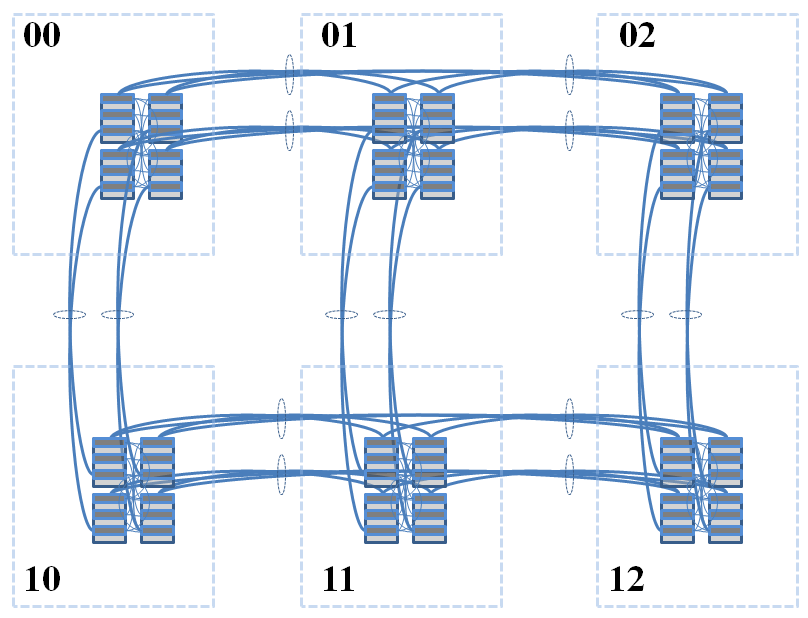}
            \label{fig_6a}
        } \\
        \subfloat[Packaging of inter-container cables between two containers into a single bundle]{
            \includegraphics[width=3.2in]{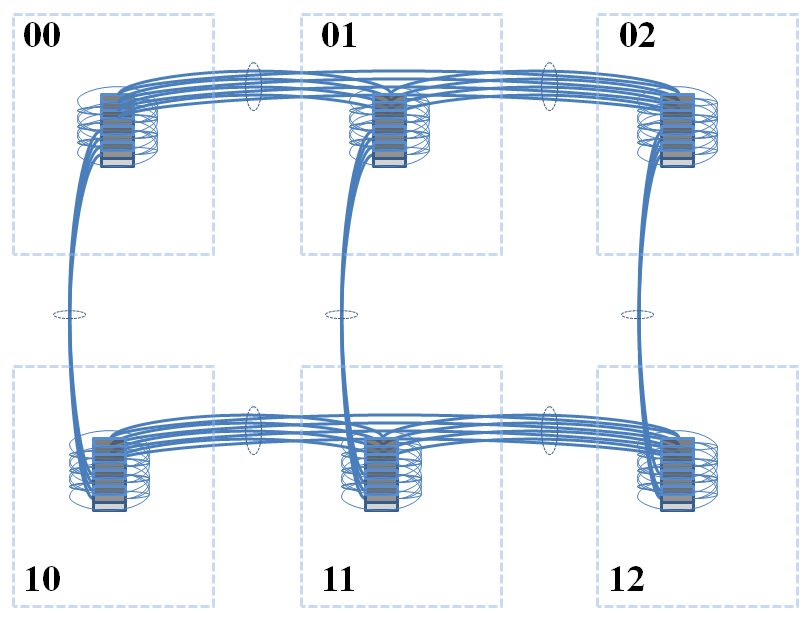}
            \label{fig_6b}
        }
        \caption{Packaging of inter-container cables}
        \captionsetup[subfigure]{font=normalsize} 
        \label{fig_6}
    \end{center}
\end{figure}

In MODRIC, the inter-container cables can also be easily bundled due to the two-dimensional arrangement of the containers. Two containers can be connected via either two bundles or a single bundle of fibers between them, as shown in Figure \ref{fig_6a} and Figure \ref{fig_6b}. Using two bundles helps in separating the EPSes into two rows and two columns based on EPSC IDs. Accommodating all four EPSCs in a single large EPSCC rack simplifies inter-container as well as intra-container connections.

\section{MODRIC NETWORKING}

MODRIC provides a modular structure with multiple equal-cost parallel paths. Due to the availability of multiple parallel paths from an access switch in MODRIC, a load-balancing routing algorithm like ECMP proves to be helpful. Decoupling the inter-container network from the intra-container network in MODRIC provides efficiency in routing. 

\subsection{Addressing}
In MODRIC, routing is done using a location-based modular L3 addressing approach. In this approach, all the VMs/servers (which may be termed as hosts) inside a container are assigned location-based IP addresses under a single class A subnet. IP addresses of the hosts inside a container follow the same IP prefix. A location-based IP address is assigned to each host dynamically in the format \(10.(k_1 \text{ bits for row ID} + k_2 \text{ bits of column ID} + k_3 \text{ bits of AS ID} + k_4 \text{ bits of host ID connected to an AS})\), where \((k_1 + k_2 + k_3 + k_4) = 24\) bits. The value of \(k_1\) and \(k_2\) determines the container ID and the IP prefix of the hosts in that container. The AS ID of an AS may be assigned in each container based on the port ID in any particular EPSC the AS is connected to. The values of \(k_1\), \(k_2\), \(k_3\), and \(k_4\) may be adjusted as per the scale of the network. Each container contains one extra server or a dedicated VM, which acts as the DHCP server responsible for assigning IP addresses to all the hosts in that container with the same IP prefix. To prevent hosts from receiving DHCP discovery messages from another DHCP server located in a different segment, inter-container link ports are disabled from forwarding DHCP packets.

This location-based modular L3 addressing in MODRIC helps locate any host within the data center and aids in routing.  

\subsection{Intra-container and inter-container routing}

As per the design objectives, routing in the inter-container network in MODRIC is decoupled from that in the intra-container network. Routing strategies to be implemented in the ASes and the EPSes are independent of each other. The EPSes will follow modular IP addressing for the inter-container routing, and the ASes will simply use the shortest path algorithm with load balancing. Modular addressing further aids in using wild-card matching of the outgoing as well as incoming packets at the EPSes for inter-container routing.

\textbf{Intra-container network:} In intra-container routing, at any AS, outgoing flows are simply forwarded to the EPSes using a multipath routing algorithm like ECMP implemented at the ASes. Incoming flows are forwarded from EPSes to the hosts using a shortest path algorithm such as Dijkstra’s algorithm. In order to avoid a loop for all the incoming flows towards the hosts, the TTL field in the IP header of the incoming packets is set to 2 at the EPSes. 

\textbf{Inter-container network:} Inter-container EPS to EPS routing follows Algorithm 1, which uses location-based IP addresses along with shortest path forwarding.

\begin{algorithm}
\caption{MODRIC Inter-Container Routing}
\textbf{Inputs}: \textit{src\_cid} (source container ID), \textit{dst\_cid} (destination container ID), \textit{curr\_cid} (current container ID where the flow currently arrives) \\
\textbf{Note}: \textit{cid} refers to the container ID, consisting of row ID and column ID.

\begin{algorithmic}[1]
\State Match the destination IP field of each flow arriving at the EPS (Edge Processing Server) at any port with the wildcard of the IP prefix.
\If{\textit{dst\_cid = curr\_cid}} \Comment{Incoming flow}
    \State Match \textit{AS\_id} with the wildcard of \textit{AS\_id} and forward the flow to the one-hop away AS.
\Else \Comment{Outgoing flow from the current container, \textit{curr\_cid}}
    \If{\textit{dst\_cid} and \textit{src\_cid} are in the same row}
        \State Forward the flow to the one-hop away EPS of the destination container via the shortest path.
    \ElsIf{\textit{dst\_cid} and \textit{src\_cid} are in the same column}
        \State Forward the flow to the one-hop away EPS of the destination container via the shortest path.
    \Else \Comment{\textit{curr\_cid = src\_cid}, and source and destination containers are not in the same row or column}
        \State Forward via a one-hop away path in either the row direction first (towards the column of the destination container) and then via a one-hop away path in the column direction (towards the row of the destination container), or vice versa.
        \State Repeat this step using ECMP (Equal-Cost Multi-Path) for each new outgoing flow toward the same destination container.
    \EndIf
\EndIf
\end{algorithmic}
\end{algorithm}

According to Algorithm 1, all the EPSes first check the IP prefix of the destination IP to determine the destination container ID (\textit{dst\_cid}) of the flow. If the flow is incoming, the destination IP of the flow is matched for the corresponding AS ID and forwarded through the port connected to that AS. For outgoing flows, if the \textit{dst\_cid} lies in the same row or column as the current container ID (\textit{curr\_cid}), then the flow is forwarded to the one-hop away EPS located in the destination container using the shortest path. If the \textit{dst\_cid} is not in the same row or column as the source container ID (\textit{src\_cid}), then ECMP (Equal-Cost Multi-Path) can be used to load balance the outgoing traffic towards the same \textit{dst\_cid} via either the row or column direction.

\subsection{OpenFlow based Networking}

The use of Software-Defined Networking (SDN) is becoming a common practice in DCNs due to the separation of the control plane and forwarding plane and the various resulting advantages. Using OpenFlow \cite{10.1145/1355734.1355746} enabled switches MODRIC makes it amenable to SDN. OpenFlow provides central control over the whole DCN either proactively or reactively. Inter-container and intra-container forwarding rules can be installed on the DCN switches by the controller. OpenFlow controller can discover topology, detect failures, and recover from failures using LLDP and BDDP protocols. Forwarding rules can be installed, updated, and removed at any time by the controller. OpenFlow improves the flexibility and manageability of the network. 

OSCAR \cite{7815328}, an OpenFlow-based DCN fabric that uses a hybrid addressing and routing scheme for modular DCNs proposed recently, can be readily adapted for MODRIC to achieve high performance as well as enhance scalability. 

\subsection{Fault-tolerant routing}
In MODRIC, failure detection is done by taking a path-level approach \cite{10.1145/2007116.2007128}, where each pair of ingress and egress EPSes has a session to monitor the path through which they are connected. In this approach, probe messages are piggybacked with the data packets so that extra probe traffic can be reduced. Further, switches do not keep any topology information; hence, extra control protocol is not necessary.

MODRIC provides fault tolerance to the routing process with the redundant links in the network. Each container has a minimum of four node-disjoint paths to another container. In case a particular inter-EPS link fails, a connected EPS sends outgoing traffic to any of the three other one-hop connected EPSCs within the container. The EPS, which receives the traffic, then forwards it via the shortest path to another container. In case an AS to EPS link fails, the AS forwards the outgoing traffic through any of the remaining working links to EPSC, and in the same case, an EPS forwards incoming traffic first to any of the other one-hop connected EPSCs within the container and then forwards the traffic to the AS. When a switch failure occurs, intra-EPSC and inter-EPSC links provide enough redundant links as backup paths to continue forwarding traffic.

\section{NETWORK COST COMPARISON}
\begin{table}[h!]
\centering
\caption{Approximate Cost per Switch Port}
\label{table:table1}
\begin{tabular}{|l|c|}
\hline
\textbf{Switch Port Capacity} & \textbf{Approx. Price per Port (US \$)} \\ \hline
1GbE                           & 100                                    \\ \hline
10GbE                          & 500                                    \\ \hline
\end{tabular}
\end{table}

\begin{table}[h!]
\centering
\caption{Approximate cost of Amphenol cables \cite{amphenol, gale2011direct}}
\label{table:table2}
\begin{tabular}{|l|c|}
\hline
\textbf{Type of Cable (Brand-Amphenol)} & \textbf{Cost (per meter)} \\ \hline
Cat 5e GbE patch cable                 & \$1.98                    \\ \hline
Cat 6 10GbE                            & \$2.60                    \\ \hline
Duplex 10 GbE MMF                      & \$12.85                   \\ \hline
\end{tabular}
\end{table}

The cost of building the DCN is an important criterion in the selection of topology for the network. In this section, we compare the cost of building MODRIC networks with that for the server-centric MDCube networks, which are meant to be low cost, and that for networks based on the popular Fattree topology. The cost of an MDC network is mainly decided by two components- the switches and the cabling. Other costs, such as the server NICs, server/ switch racks, etc., can be ignored as these are relatively less significant. 

We shall derive the expressions for the volume of switches and the cables required for the three topologies. For the switches, the expressions shall be derived in terms of the number of switching ports required, and they shall be the lengths of cables required for the cabling to simplify the cost estimation. To get the overall costs of construction for MDCs of similar size to the considered topologies, typical prevailing market prices for the switching ports and cables shall be used. 

It is common for the DCNs to use commodity devices to achieve lower costs. Several Ethernet top-of-rack (TOR) switches of GbE or 10GbE capacity are becoming commodities \cite{88}. For example, two types of TOR switches- TOR1G and TOR10G are being widely used. A TOR1G commodity switch is a 48-port GbE switch with 4 additional 10GbE ports. A TOR10G switch has 24-port/48-port 10GbE ports. In the TOR1G switch, the 4 x 10GbE ports support SFP (small form-factor pluggable) \cite{chen2019small}, providing options for both copper and fiber optic connectivity. The 10GbE Ethernet RJ45 ports support 10GBASE-T copper cables of lengths up to 100m. A multimode fiber (MMF) cable allows 10GbE connections with a range of more than 300m. 

In a TOR10G switch, commonly, there are 4 x 10GbE ports with SFP and the remaining 10GbE RJ45 ports. TOR10G switches consisting of either 24 x 10GbE SFP ports or 48 x 10GbE SFP ports are also available, where configuration may vary slightly based on the model. Table \ref{table:table1} shows the per port switch costs for conventional ASIC-based commodity TOR switches as obtained from \cite{mudigonda2011taming, arista}.

Considering the link speed and length requirement, three variants of cables are needed, as listed in Table \ref{table:table2}, along with their approximate costs.

\subsection{Cost of Switches}
For the evaluation of switch costs for MODRIC, we will use Property 5 in Section III, which gives the number of switch ports for the network. For example, a 10x10 MODRIC network can have \( m \cdot n \cdot (m+n-2) \cdot \left( \frac{4C}{c} \right) = 72000 \) hosts when link speeds of \( C = 10 \, \text{Gbps} \) and \( c = 1 \, \text{Gbps} \) are used. The network can be constructed with \( q = 2 \) number of 24-port TOR10G EPSes in each EPSC, \( m+n-2 = 18 \) TOR1G ASes connected to the EPSCs, and \( \frac{4C}{c} = 40 \) hosts connected to each AS. Each container will have 720 hosts. The total number of 10GbE switch ports required will be \( 4 \cdot m \cdot n \cdot \left( q^2 + 2q + 3m + 3n - 6 \right) = 24800 \), and the number of GbE switch ports required will be \( m \cdot n \cdot (m+n-2) \cdot \left( \frac{4C}{c} \right) = 72000 \).

If 6U racks are used for an EPSC, one can have \( q \) up to 6. With 24-port TOR10GbE EPSes, it is possible to construct a 25x25 MODRIC network that can have 1,200,000 hosts.

However, a MODRIC MDC can also be built using 48-port TOR10G switches as EPSes placed inside EPSCs built up of 12U racks. With this configuration, MODRIC can house a maximum of 103x103 containers with 103,883,328 servers.

A similar MDCube \cite{10.1145/1658939.1658943} topology is taken for comparison. An MDCube uses only TOR1G switches and servers with multiple Ethernet ports for its construction. A 2-D MDCube topology is constructed with \( r \)-port switches and BCubek with \( k+1 \) levels of switches. An MDCube topology with such configuration consists of a total of \( rk \cdot (k+1) \) switches and \( rk+1 \) hosts per BCube container. Thus, there exist \( r \cdot rk \cdot (k+1) = rk+1 \cdot (k+1) \) GbE ports and \( z \cdot rk \cdot (k+1) \) 10GbE ports per container, where \( z \) is the number of 10GbE ports per switch. However, for an \( m \times n \) 2D MDCube, \( z \cdot (m+n-2) \) numbers of 10GbE ports per container are actually used for inter-container connectivity. The value of \( k \) is taken as 1 to minimize the number of switches and server ports for cost-effectiveness.

The Fat tree topology \cite{88} is also taken for comparison, which is a switch-centric topology built from TOR1G commodity switches. It maintains equal radix at all the switches distributed in three layers. A Fattree topology with \( R \)-port switches contains \( R^3/4 \) servers and \( 5R^2/4 \) switches. Thus, the total number of GbE ports used in the topology becomes \( \left( \frac{5R^2}{4} \right) \cdot R = \frac{5R^3}{4} \).

\begin{table*}[t]
\renewcommand{\arraystretch}{1.2} 
\setlength{\tabcolsep}{3pt} 
\centering
\caption{No. of Switch Ports in MODRIC, MDCube and Fattree}
\begin{tabular}{|p{2.7cm}|p{2cm}|p{2cm}|p{2cm}|p{2cm}|p{2cm}|p{2cm}|}
\hline
\textbf{Dimensions/ Configuration} & \textbf{No. of Hosts} & \multicolumn{3}{c|}{\textbf{Estimated Switch Cost}} & \textbf{Cost of Cables (M\$)} & \textbf{Total Cost of Switches \& Cables (M\$)} \\ \cline{3-5}
 &  & \textbf{No. of 10GbE Ports} & \textbf{No. of GbE Ports} & \textbf{Total Switch Cost (M\$)} &  &  \\ \hline
\multicolumn{7}{|c|}{\textbf{MODRIC}} \\ \hline
 & $m.n.(m+n-2).(4C/c)$ & $4.m.n.(q^2 +2q+3m+3n-6)$ & $m.n.(m+n-2).(4C/c)$ & -- & -- & -- \\ \hline
$m=10, n=10, q=2, x=18, c=1GbE, C=10GbE$ & 72000 & 24800 & 72000 & 19.6 & 1.136 & 20.736 \\ \hline
$m=15, n=15, q=3, x=28, c=1GbE, C=10GbE$ & 252000 & 89100 & 252000 & 69.75 & 5.345 & 75.095 \\ \hline
$m=20, n=20, q=5, x=38, c=1GbE, C=10GbE$ & 608000 & 238400 & 608000 & 180 & 16.264 & 196.264 \\ \hline
$m=25, n=25, q=6, x=48, c=1GbE, C=10GbE$ & 1200000 & 480000 & 1200000 & 360 & 38.799 & 398.799 \\ \hline
\multicolumn{7}{|c|}{\textbf{MDCube}} \\ \hline
 & $m.n.r^{k+1}$ & $m.n.z.(m+n-2)$ & $m.n.r^{k+1}.(k+1)$ & -- & -- & -- \\ \hline
$m=10, n=10, r=27, k=1, z=4$ & 72900 & 7200 & 145800 & 18.18 & 1.471 & 19.651 \\ \hline
$m=15, n=15, r=33, k=1, z=4$ & 245025 & 25200 & 490050 & 61.605 & 6.491 & 68.096 \\ \hline
$m=20, n=20, r=39, k=1, z=4$ & 608400 & 60800 & 1216800 & 152.08 & 19.229 & 171.309 \\ \hline
$m=25, n=25, r=44, k=1, z=4$ & 1210000 & 120000 & 2420000 & 302 & 44.797 & 346.797 \\ \hline
\multicolumn{7}{|c|}{\textbf{Fat Tree}} \\ \hline
 & $R^3/4$ & -- & $5R^3/4$ & -- & -- & -- \\ \hline
$R=66$ & 71874 & Nil & 359370 & 35.937 & 0.571 & 35.508 \\ \hline
$R=100$ & 250000 & Nil & 1250000 & 125 & 1.985 & 126.985 \\ \hline
$R=134$ & 601526 & Nil & 3007630 & 300.763 & 4.773 & 305.546 \\ \hline
$R=168$ & 1185408 & Nil & 5927040 & 592.704 & 9.402 & 602.106 \\ \hline
\end{tabular}
\label{table:table3}
\end{table*}

The estimated costs of switches of all three topologies for different network sizes are shown in Table \ref{table:table3}.

\subsection{Cost of Cabling}

\begin{figure}[!t]
        \begin{center}
            \includegraphics[width=3.2in]{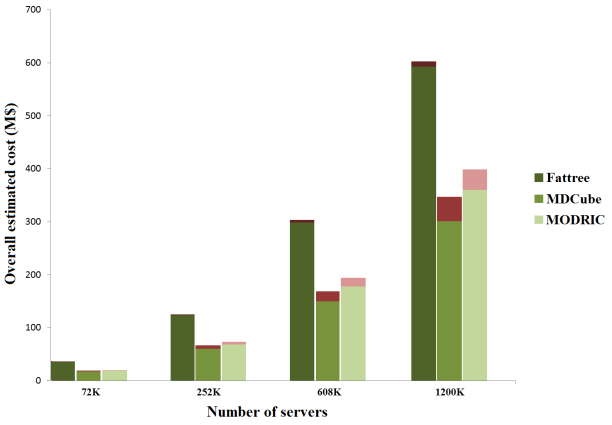}
            \caption{Comparison of overall estimated cost among fattree, MDCube and MODRIC (Lower parts of the bars show the cost of switches and the upper parts show the cost of cabling)}
            \label{fig_7}
        \end{center}
    \end{figure}

In MODRIC, $6U$ or $12U$ racks can be suitably used to place TOR10G EPSes to form EPSCs, and $42U$ racks can be used to place the TOR1G ASes within the container. Conventional $20 \, \text{ft}$ or $40 \, \text{ft}$ shipping containers [\cite{wikipedia_contributors_2019}] can be used for the DCN containers. For comparison, $10 \, \text{GbE}$ fiber optic cables are used for inter-container links in both MODRIC and MDCube. For intra-container links, MODRIC uses $10 \, \text{GbE}$ copper between any two switches and $1 \, \text{GbE}$ links between switch and server, whereas MDCube uses only $1 \, \text{GbE}$ links. In fat tree topology, only $1 \, \text{GbE}$ copper links are used as per the design proposed. For the sake of simplicity in comparison, only two types of average cable lengths are taken such as - $1 \, \text{m}$ and $3 \, \text{m}$ cable in all the topologies. We assume that intra-rack or intra-pod cables can be limited to an average length of $1 \, \text{m}$ and those used for inter-rack or inter-pod connectivity can be limited to an average length of $3 \, \text{m}$. However, inter-container connections in MODRIC and MDCube require longer and higher speed cables which can be satisfied by optical fibers only. For comparison between MODRIC and MDCube, inter-container cabling structure in both are kept almost same as both these topologies use four $10 \, \text{GbE}$ links to connect any two containers located in the same row or column. Thus, total length of inter-container cables required connecting an $m \times n$ MODRIC or an MDCube with same dimensions can be computed based on Property 6. For the purpose of evaluation, consider a conventional $20 \, \text{foot}$ shipping container the DCN for both MODRIC and MDCube, which has the dimensions of length, $L = 6.1 \, \text{m}$ and width, $W = 2.44 \, \text{m}$ [\cite{wikipedia_contributors_2019}]. Assume further that any two adjacent containers in the same row or column are located $1 \, \text{m}$ away from each other i.e., $d_L = d_W = 1 \, \text{m}$.

A MODRIC container requires $6q \times 10 \, \text{GbE}$ copper cables for connecting $q$ layers of inter-EPSC meshes. It requires $\frac{4q(q - 1)}{2} = 2q(q - 1) \times 10 \, \text{GbE}$ copper links for intra-EPSC connections in four EPSCs. $4x \times 10 \, \text{GbE}$ links are used to connect $x$ ASes to EPSCs. $x \times y \times \text{GbE}$ links are used for connecting hosts to ASes.

An MDCube container uses $rk \cdot r = rk+1 \times \text{GbE}$ links to connect hosts to the layer-0 TOR switches placed in their corresponding racks. The number of GbE links required to connect layer-$k$ switches to the hosts is --
\[
\sum_{k=1}^{K} k \cdot r^{k} \cdot r = \sum_{k=1}^{K} k \cdot r^{k+1}
\]

Inter-container cabling length in MDCube is also evaluated using property 6 of MODRIC.

A fattree topology with radix $R$ switches uses $\left(\frac{R^3}{4}\right) \times \text{GbE}$ links to connect access switches to the hosts, $\left(\frac{R^2}{4}\right) \times \text{GbE}$ links to connect access switches to aggregation switches and $\left(\frac{R^3}{4}\right) \times \text{GbE}$ links to connect aggregation links to core switches.

The total estimated costs of cabling of the networks, based on the unit costs in Table \ref{table:table2}, are shown in Table \ref{table:table3}. 

The overall estimated costs of the networks are also shown in Table \ref{table:table3} and the bar chart in Figure \ref{fig_7}. It can be observed that the MDCube networks have a marginally lower cost than MODRIC. The Fattree networks have the highest total cost. The cabling cost is, however, lowest for Fattree-based networks as these use only $1$GbE copper links.

Merchant silicon-based networking ASICs are also now becoming available, which provide high energy efficiency and cost advantage \cite{1676437}. Several makers have come up with such low-cost ASICs. As an example, a Fujitsu MB86C69RBC switch \cite{alldatasheet_com_2025} ASIC costs around \$410 and consumes very low power of around 22W only \cite{5238680}. Such an ASIC board requires integration with a CPU and per port PHY and SFP+ chips in order to build the switch. The MB86C69RBC switch chip supports 26 x 10GbE ports and two additional GbE management ports \cite{alldatasheet_com_2025}. Summing these up, in Table \ref{table:table4}, the estimated cost for a merchant silicon ASIC-based 10GbE switch is \$280 per port. The result is a substantial reduction in the cost of MODRIC networks (Table \ref{table:table5}) that will probably erase any cost advantage of server-centric networks.

\begin{table}[h!]
\centering
\caption{Estimating cost of the merchant silicon ASIC-based switch}
\begin{tabular}{|l|c|c|c|c|c|}
\hline
\textbf{Switch Part} & \textbf{ASIC} & \textbf{CPU} & \textbf{PHY} & \textbf{SFP+} & \textbf{Total} \\
\hline
\textbf{Cost (\$)} & 410/26 $\approx$ 15 & 130/26 = 5 & 10 & 250 & 280 \\
\hline
\end{tabular}
\label{table:table4}
\end{table}

\section{PERFORMANCE EVALUATION}
The performance of MODRIC is tested and compared with DCN interconnection designs such as fattree \cite{88}, Jellyfish \cite{180604}, and MDC designs such as MDCube and uFix in an SDN environment using Mininet \cite{10.1145/1868447.1868466}. Mininet provides a real virtual network with OVS software switches that support SDN API, such as OpenFlow. Floodlight OpenFlow controller \cite{floodlight} is used to install architecture-specific routing mechanisms in the topologies. The Floodlight controller runs in an Ubuntu VM allocated with 4GB RAM and 10 CPU cores. The Ubuntu VM is mounted on an IBM x3500 M4 server with Intel(R) Xeon(R) E5-2620 2.00GHz processor. Mininet VM is allocated 10GB RAM and 4 CPU cores mounted on another machine with an Intel i7 3.40GHz processor.

For the experiments, networks with the following topologies are created in Mininet -- a fat tree with radix 8 is taken with 128 hosts, a Jellyfish with 128 hosts, a 2x4 MDCube with 2x4 BCube(4,1) containers, a level-2 uFix domain containing a total of 144 hosts with 3 level-1 uFix domains each having 3 BCube(4,1) containers and a 2x4 MODRIC with containers each having 1 EPS per EPSC, 4 ASes and 16 hosts. In each network, a group of 16 hosts represents a container in an MDC. For example, a pod in fattree topology and a random cluster of servers in Jellyfish form a group. In uFix, any 8 groups with 128 hosts out of the total of 9 groups with 144 hosts in the network are taken for fair comparison. In all the topologies, switch-to-host links are kept at 100 Mbps, and a switch-to-switch inter-container link is kept at 400 Mbps. 

The performances of the topologies are tested against different traffic conditions. D-ITG traffic generator \cite{BOTTA20123531} is used to generate different traffic patterns with all the hosts creating flows with 100Mbps data rate each inside the topologies.

All the experiments are repeated 20 times for each given parameter, and the aggregated results are averaged to compute the final result.

\subsection{Experiment Set-1 –- Performance comparison against traffic load:}

\begin{table}[t]
\renewcommand{\arraystretch}{1.2} 
\setlength{\tabcolsep}{2pt} 
\centering
\caption{Estimated costs of MODRIC networks built using Merchant Silicon ASIC-based TOR10G switches}
\begin{tabular}{|p{1.2cm}|p{1.2cm}|p{1.2cm}|p{1.2cm}|p{1.2cm}|p{1.4cm}|}
\hline
\multirow{2}{*}{\textbf{\begin{tabular}[c]{@{}c@{}}Network \\Size (No. \\ of Hosts)\end{tabular}}} & \multicolumn{3}{c|}{\textbf{Estimating Switch Cost}}                                                    & \multirow{2}{*}{\textbf{\begin{tabular}[c]{@{}c@{}}Cost of \\ Cables \\ (M\$)\end{tabular}}} & \multirow{2}{*}{\textbf{\begin{tabular}[c]{@{}c@{}}Total Cost \\ of Switches \\ \& Cables \\ (M\$)\end{tabular}}} \\ \cline{2-4}
 & \textbf{\begin{tabular}[c]{@{}c@{}}No. of \\ 10GbE \\ Ports\end{tabular}} & \textbf{\begin{tabular}[c]{@{}c@{}}No. of \\ GbE \\ Ports\end{tabular}} & \textbf{\begin{tabular}[c]{@{}c@{}}Total \\Switch \\ Cost (M\$)\end{tabular}} &  &  \\ \hline
72000 & 24800 & 72000 & 14.144 & 1.136 & 15.280 \\ \hline
252000 & 89100 & 252000 & 50.150 & 5.345 & 55.496 \\ \hline
608000 & 238400 & 608000 & 127.550 & 16.264 & 144.814 \\ \hline
1200000 & 480000 & 1200000 & 254.400 & 38.799 & 293.199 \\ \hline
\end{tabular}
\label{table:table5}
\end{table}

\begin{figure}[!t]
        \begin{center}
            \includegraphics[width=3.2in]{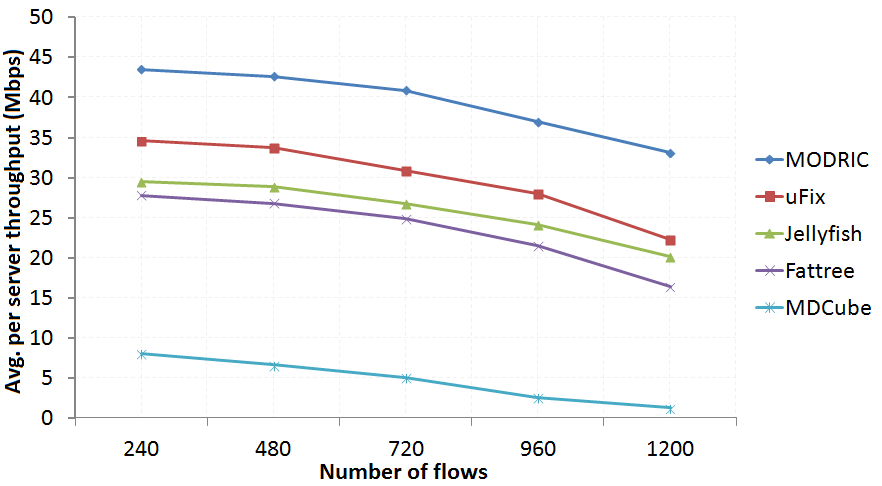}
            \caption{Average per-server throughput under increasing traffic load}
            \label{fig_8}
        \end{center}
    \end{figure}

In this experiment, 15 out of 16 hosts in each group send TCP traffic, and the remaining host is kept as a log server for D-ITG. Each host creates at least \( n \) TCP connections with \( N \) other hosts chosen randomly located in 8 groups, including its own with \( N = 2, 4, 6, 8, 10 \). Each of the \( N \) TCP flows lasts for 100 seconds. The resulting average per-server throughput is shown in Figure \ref{fig_8}. MODRIC outperforms the other topologies for increasing traffic load.

\subsection{Experiment Set-2 –- Performance with traffic generated by all hosts: }

\begin{figure}[!t]
        \begin{center}
            \includegraphics[width=3.2in]{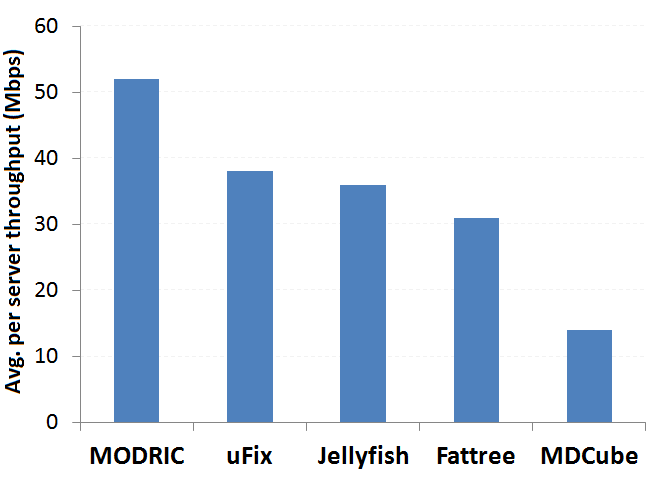}
            \caption{Average per-server throughput comparison under a traffic pattern where each host sends a traffic flow to another host}
            \label{fig_9}
        \end{center}
    \end{figure}

In this experiment, every host sends and receives 100Mbps traffic to and from another host chosen randomly over two TCP connections. No host communicates over more than two TCP connections. For 128 hosts, 64 unique pairs of hosts communicate over a total of 128 TCP connections. In all the topologies, 7 pairs of hosts in one group send traffic to 7 different pairs of hosts located in 7 other groups, and a single pair of hosts within a group communicate. In this experiment, an extra log server is used for each group.

MODRIC shows better throughput than the others [Figure \ref{fig_9}]. Performances of uFix, Jellyfish, and fat tree are comparable, whereas MDCube shows the lowest throughput among all. 

\subsection{Experiment Set-3 –- Container to container one-to-one and all-to-all traffic performance comparison:}

\begin{figure}[!t]
        \begin{center}
            \includegraphics[width=3.2in]{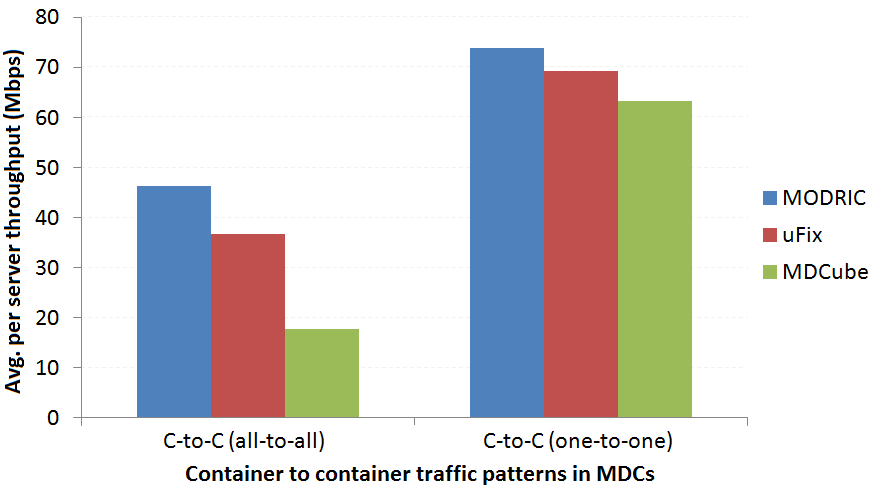}
            \caption{ Average per-server throughput under container-to-container traffic shown by MDC topologies}
            \label{fig_10}
        \end{center}
    \end{figure}

In order to evaluate container to container traffic capacity of MODRIC, we perform two experiments for bandwidth-intensive traffic typical of many present-day applications \cite{apache_software_foundation_2019, dean2004mapreduce}. The first scenario is for one-to-one traffic between a pair of containers where each of the 16 hosts in a source container sets up a single TCP connection with one unique host in the destination container. A total of 16 flows, each of 100 Mbps, between the two containers are passed for 100 seconds, and the average throughputs are observed. The second scenario is for all-to-all traffic, where each of the 16 hosts in one container sets up one TCP connection with each of the 16 hosts in another container. Thus, a total of \( 16 \times 16 = 256 \) TCP flows, each of 100 Mbps, are transmitted between two containers for 100 seconds, and the average throughput per server is observed. The two diagonally opposite containers in the network are taken for the experiments in MODRIC and MDCube. For uFix, two containers in two different level-1 domains are taken. The performance of the MDC topologies is shown in Figure \ref{fig_10}. MODRIC shows higher per-server throughput than the other two in both cases. The higher inter-container throughput performance of MODRIC is due to the good number of node disjoint paths. The difference in performance is more pronounced for the C-to-C (all-to-all) traffic scenario.

\subsection{Experiment Set-4 -- Performance of MODRIC under failures}

\begin{figure}[!t]
        \begin{center}
            \includegraphics[width=3.2in]{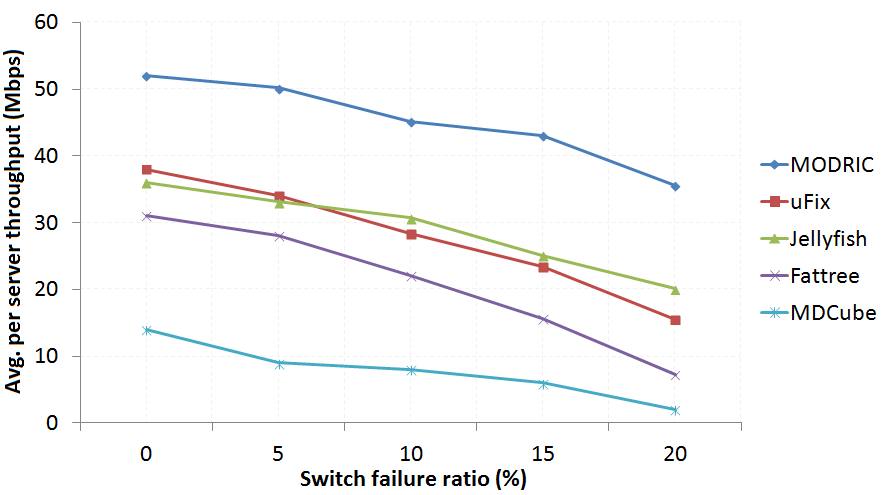}
            \caption{Average per-server throughput against switch failures under traffic communicated by all hosts}
            \label{fig_11}
        \end{center}
    \end{figure}

MODRIC is switch-centric topology, major failures are contributed by the switches only and server failures are not significant. In this experiment, we evaluate the fault tolerance performance of MODRIC in terms of switch failures only. In order to compare the performances of all the topologies under switch failures, we generate the same traffic pattern as shown in experiment 2. Figure \ref{fig_11} shows the performances of all the topologies under an increasing percentage of switch failures. MODRIC outperforms other topologies significantly. Jellyfish outperforms uFix in terms of a larger switch failure ratio. Jellyfish and uFix show higher fault tolerance than fat-tree, and MDCube and fat-tree show better results than MDCube. 

\subsection{Experiment Set-5 –- Routing performance of MODRIC }

\begin{figure}[!t]
        \begin{center}
            \includegraphics[width=3.2in]{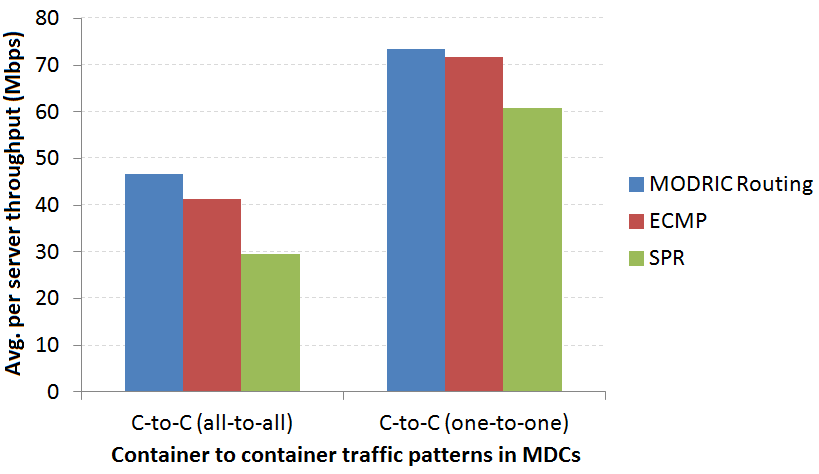}
            \caption{Routing scheme performance comparison under container-to-container traffic pattern}
            \label{fig_12}
        \end{center}
    \end{figure}

\begin{figure}[!t]
        \begin{center}
            \includegraphics[width=3.2in]{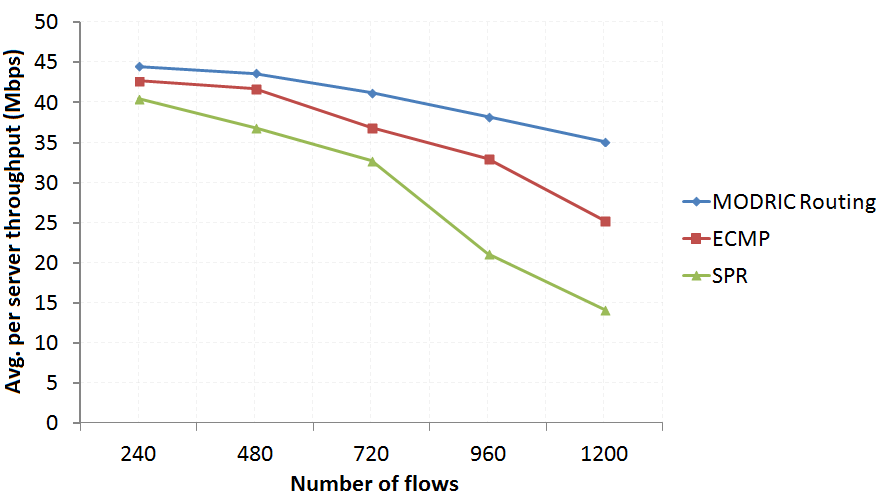}
            \caption{Routing scheme performance comparison under increasing traffic load}
            \label{fig_13}
        \end{center}
    \end{figure}

We implemented MODRIC routing, ECMP only, and Single Path Routing (SPR) using Floodlight controller and applied it to MODRIC topology. In ECMP only, ECMP is used between pairs of ASes without using any addressing scheme and without separating inter-container and intra-container routing. In Single Path Routing, traffic is forwarded via only a single shortest path. The performances of MODRIC are observed for all three mechanisms. The network in each case is tested under two scenarios: 1. using the container-to-container traffic pattern of experiment 3 and 2. using the traffic pattern of experiment 1.  MODRIC routing scheme shows better performance than ECMP, and both of these schemes show significantly better results than SPR in scenario 1 [Figure \ref{fig_12}]. MODRIC routing schemes give a better, steady performance with increasing traffic load, whereas performance degrades for the ECMP-only approach due to the high computation requirement [Figure \ref{fig_13}]. SPR underperforms the other two schemes due to congestion at some links.

\subsection{Experiment Set-6 –- Performance of MODRIC with an increasing number of containers}

\begin{figure}[!t]
        \begin{center}
            \includegraphics[width=3.2in]{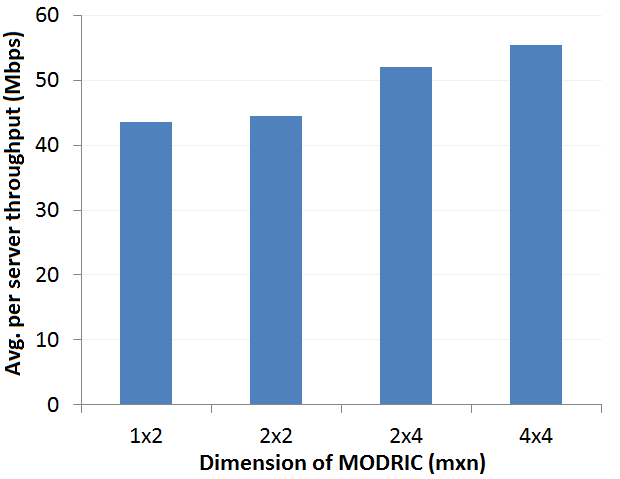}
            \caption{Throughput performance of MODRIC with increasing dimensions (grids)}
            \label{fig_14}
        \end{center}
    \end{figure}

MODRIC network can be expanded by increasing the number of containers in it, which in turn increases the capacity of the network. In this experiment, we test the effect of increasing container numbers on the network capacity. MODRIC network with 128 hosts is arranged in four different grid formations with four different dimensions. The same 128 hosts are shared using a total of 2, 4, 8, and 16 containers, respectively. Figure \ref{fig_14} shows the improvement in network throughput with an increasing number of containers in the network.

\section{CONCLUSION}

In this paper, we have proposed MODRIC, a new shipping container-based switch-centric modular data center network topology using commodity switches. The proposal includes the design for the inter-container network that is similar to the generalized hypercube and a flexible design for the container network. The resulting overall network has been shown to possess excellent topological properties, including low constant diameter, a good number of parallel paths, high bi-section bandwidth, and high scalability. The cabling schemes proposed indicate an easy layout of the network. Building cost comparisons with representative topologies presented are favorable to the proposed topology. The paper presents suitable addressing and routing schemes that take advantage of the topology. Existing OpenFlow-based hybrid addressing and routing schemes for modular data center networks can also be readily adapted for high-performance topology. The performance evaluation presented shows that the proposed topology performs significantly better than existing topologies under different traffic conditions in terms of network throughput. MODRIC has also been shown to have much better fault tolerance.

\bibliographystyle{IEEEtran}
 	\bibliography{sample}

\begin{thebibliography}{10}
\providecommand{\url}[1]{#1}
\csname url@samestyle\endcsname
\providecommand{\newblock}{\relax}
\providecommand{\bibinfo}[2]{#2}
\providecommand{\BIBentrySTDinterwordspacing}{\spaceskip=0pt\relax}
\providecommand{\BIBentryALTinterwordstretchfactor}{4}
\providecommand{\BIBentryALTinterwordspacing}{\spaceskip=\fontdimen2\font plus
\BIBentryALTinterwordstretchfactor\fontdimen3\font minus \fontdimen4\font\relax}
\providecommand{\BIBforeignlanguage}[2]{{%
\expandafter\ifx\csname l@#1\endcsname\relax
\typeout{** WARNING: IEEEtran.bst: No hyphenation pattern has been}%
\typeout{** loaded for the language `#1'. Using the pattern for}%
\typeout{** the default language instead.}%
\else
\language=\csname l@#1\endcsname
\fi
#2}}
\providecommand{\BIBdecl}{\relax}
\BIBdecl

\bibitem{9925714}
M.~Lv, J.~Fan, W.~Fan, and X.~Jia, ``A high-performantal and server-centric based data center network,'' \emph{IEEE Transactions on Network Science and Engineering}, vol.~10, no.~2, pp. 592--605, 2023.

\bibitem{apache_software_foundation_2019}
\BIBentryALTinterwordspacing
A.~S. Foundation, ``Apache hadoop,'' 2019. [Online]. Available: \url{https://hadoop.apache.org/}
\BIBentrySTDinterwordspacing

\bibitem{dean2004mapreduce}
J.~Dean and S.~Ghemawat, ``Mapreduce: Simplified data processing on large clusters,'' 2004.

\bibitem{varia_mathew_2014}
\BIBentryALTinterwordspacing
J.~Varia and S.~Mathew, \emph{Amazon Web Services -Overview of Amazon Web Services}, 2014. [Online]. Available: \url{https://media.amazonwebservices.com/AWS_Overview.pdf}
\BIBentrySTDinterwordspacing

\bibitem{10535462}
G.~Wang, Y.~Zhang, J.~Yu, M.~Ma, C.~Hu, J.~Fan, and L.~Zhang, ``Hs-dcell: A highly scalable dcell-based server-centric topology for data center networks,'' \emph{IEEE/ACM Transactions on Networking}, vol.~32, no.~5, pp. 3808--3823, 2024.

\bibitem{jia2023sretor}
Z.~Jia, Q.~Liu, and Y.~Sun, ``sretor: a semi-centralized regular topology routing scheme for data center networking,'' \emph{Journal of Cloud Computing}, vol.~12, no.~1, p. 150, 2023.

\bibitem{sun}
\BIBentryALTinterwordspacing
\emph{Sun Modular Datacenter S20/D20 Overview}, Sun Microsystems, Santa Clara, California, USA, 2008. [Online]. Available: \url{https://docs.oracle.com/cd/E19115-01/mod.dc.s20/820-5770-10/820-5770-10.pdf}
\BIBentrySTDinterwordspacing

\bibitem{kotsis1992interconnection}
G.~Kotsis, \emph{Interconnection topologies and routing for parallel processing systems}.\hskip 1em plus 0.5em minus 0.4em\relax ACPC-Austrian Center for Parallel Computation, 1992.

\bibitem{88}
\BIBentryALTinterwordspacing
M.~Al-Fares, A.~Loukissas, and A.~Vahdat, ``A scalable, commodity data center network architecture,'' in \emph{Proceedings of the ACM SIGCOMM 2008 Conference on Data Communication}, ser. SIGCOMM '08.\hskip 1em plus 0.5em minus 0.4em\relax New York, NY, USA: Association for Computing Machinery, 2008, p. 63–74. [Online]. Available: \url{https://doi.org/10.1145/1402958.1402967}
\BIBentrySTDinterwordspacing

\bibitem{https://doi.org/10.1002/j.1538-7305.1953.tb01433.x}
\BIBentryALTinterwordspacing
C.~Clos, ``A study of non-blocking switching networks,'' \emph{Bell System Technical Journal}, vol.~32, no.~2, pp. 406--424, 1953. [Online]. Available: \url{https://onlinelibrary.wiley.com/doi/abs/10.1002/j.1538-7305.1953.tb01433.x}
\BIBentrySTDinterwordspacing

\bibitem{180604}
\BIBentryALTinterwordspacing
A.~Singla, C.-Y. Hong, L.~Popa, and P.~B. Godfrey, ``Jellyfish: Networking data centers randomly,'' in \emph{9th USENIX Symposium on Networked Systems Design and Implementation (NSDI 12)}.\hskip 1em plus 0.5em minus 0.4em\relax San Jose, CA: USENIX Association, Apr. 2012, pp. 225--238. [Online]. Available: \url{https://www.usenix.org/conference/nsdi12/technical-sessions/presentation/singla}
\BIBentrySTDinterwordspacing

\bibitem{guo2009bcube}
C.~Guo, G.~Lu, D.~Li, H.~Wu, X.~Zhang, Y.~Shi, C.~Tian, Y.~Zhang, and S.~Lu, ``Bcube: a high performance, server-centric network architecture for modular data centers,'' in \emph{Proceedings of the ACM SIGCOMM 2009 conference on Data communication}, 2009, pp. 63--74.

\bibitem{10.1145/1658939.1658943}
\BIBentryALTinterwordspacing
H.~Wu, G.~Lu, D.~Li, C.~Guo, and Y.~Zhang, ``Mdcube: a high performance network structure for modular data center interconnection,'' in \emph{Proceedings of the 5th International Conference on Emerging Networking Experiments and Technologies}, ser. CoNEXT '09.\hskip 1em plus 0.5em minus 0.4em\relax New York, NY, USA: Association for Computing Machinery, 2009, p. 25–36. [Online]. Available: \url{https://doi.org/10.1145/1658939.1658943}
\BIBentrySTDinterwordspacing

\bibitem{1676437}
Bhuyan and Agrawal, ``Generalized hypercube and hyperbus structures for a computer network,'' \emph{IEEE Transactions on Computers}, vol. C-33, no.~4, pp. 323--333, 1984.

\bibitem{6089059}
D.~Li, M.~Xu, H.~Zhao, and X.~Fu, ``Building mega data center from heterogeneous containers,'' in \emph{2011 19th IEEE International Conference on Network Protocols}, 2011, pp. 256--265.

\bibitem{10.1145/1921168.1921189}
\BIBentryALTinterwordspacing
L.~Popa, S.~Ratnasamy, G.~Iannaccone, A.~Krishnamurthy, and I.~Stoica, ``A cost comparison of datacenter network architectures,'' in \emph{Proceedings of the 6th International COnference}, ser. Co-NEXT '10.\hskip 1em plus 0.5em minus 0.4em\relax New York, NY, USA: Association for Computing Machinery, 2010. [Online]. Available: \url{https://doi.org/10.1145/1921168.1921189}
\BIBentrySTDinterwordspacing

\bibitem{10.1145/1250662.1250679}
\BIBentryALTinterwordspacing
J.~Kim, W.~J. Dally, and D.~Abts, ``Flattened butterfly: a cost-efficient topology for high-radix networks,'' in \emph{Proceedings of the 34th Annual International Symposium on Computer Architecture}, ser. ISCA '07.\hskip 1em plus 0.5em minus 0.4em\relax New York, NY, USA: Association for Computing Machinery, 2007, p. 126–137. [Online]. Available: \url{https://doi.org/10.1145/1250662.1250679}
\BIBentrySTDinterwordspacing

\bibitem{5238680}
N.~Farrington, E.~Rubow, and A.~Vahdat, ``Data center switch architecture in the age of merchant silicon,'' in \emph{2009 17th IEEE Symposium on High Performance Interconnects}, 2009, pp. 93--102.

\bibitem{10.1145/1355734.1355746}
\BIBentryALTinterwordspacing
N.~McKeown, T.~Anderson, H.~Balakrishnan, G.~Parulkar, L.~Peterson, J.~Rexford, S.~Shenker, and J.~Turner, ``Openflow: enabling innovation in campus networks,'' \emph{SIGCOMM Comput. Commun. Rev.}, vol.~38, no.~2, p. 69–74, Mar. 2008. [Online]. Available: \url{https://doi.org/10.1145/1355734.1355746}
\BIBentrySTDinterwordspacing

\bibitem{7815328}
N.~Medhi and D.~K. Saikia, ``Openflow-based scalable routing with hybrid addressing in data center networks,'' \emph{IEEE Communications Letters}, vol.~21, no.~5, pp. 1047--1050, 2017.

\bibitem{10.1145/2007116.2007128}
\BIBentryALTinterwordspacing
M.~Suchara, D.~Xu, R.~Doverspike, D.~Johnson, and J.~Rexford, ``Network architecture for joint failure recovery and traffic engineering,'' \emph{SIGMETRICS Perform. Eval. Rev.}, vol.~39, no.~1, p. 97–108, Jun. 2011. [Online]. Available: \url{https://doi.org/10.1145/2007116.2007128}
\BIBentrySTDinterwordspacing

\bibitem{amphenol}
\BIBentryALTinterwordspacing
2024. [Online]. Available: \url{http://www.cablesondemand.com/}
\BIBentrySTDinterwordspacing

\bibitem{gale2011direct}
R.~Gale, ``Direct attach cable assembly specification analysis from 8 port s-parameters,'' Ph.D. dissertation, Texas Tech University, 2011.

\bibitem{chen2019small}
Y.-M. Chen, ``Small form-factor pluggable transceiver,'' Jan.~29 2019, uS Patent 10,193,590.

\bibitem{mudigonda2011taming}
J.~Mudigonda, P.~Yalagandula, and J.~C. Mogul, ``Taming the flying cable monster: A topology design and optimization framework for $\{$Data-Center$\}$ networks,'' in \emph{2011 USENIX Annual Technical Conference (USENIX ATC 11)}, 2011.

\bibitem{arista}
\BIBentryALTinterwordspacing
 [Online]. Available: \url{https://www.arista.com/assets/data/pdf/10GigE_Whitepaper.pdf}
\BIBentrySTDinterwordspacing

\bibitem{wikipedia_contributors_2019}
\BIBentryALTinterwordspacing
W.~Contributors, ``Intermodal container,'' Nov 2019. [Online]. Available: \url{https://en.wikipedia.org/wiki/Intermodal_container}
\BIBentrySTDinterwordspacing

\bibitem{alldatasheet_com_2025}
\BIBentryALTinterwordspacing
alldatasheet.com, ``Mb86c69rbc pdf,'' 2025. [Online]. Available: \url{http://pdf1.alldatasheet.com/datasheet-pdf/view/421662/FUJITSU/MB86C69RBC.html}
\BIBentrySTDinterwordspacing

\bibitem{10.1145/1868447.1868466}
\BIBentryALTinterwordspacing
B.~Lantz, B.~Heller, and N.~McKeown, ``A network in a laptop: rapid prototyping for software-defined networks,'' in \emph{Proceedings of the 9th ACM SIGCOMM Workshop on Hot Topics in Networks}, ser. Hotnets-IX.\hskip 1em plus 0.5em minus 0.4em\relax New York, NY, USA: Association for Computing Machinery, 2010. [Online]. Available: \url{https://doi.org/10.1145/1868447.1868466}
\BIBentrySTDinterwordspacing

\bibitem{floodlight}
\BIBentryALTinterwordspacing
2025. [Online]. Available: \url{https://groups.io/g/floodlight}
\BIBentrySTDinterwordspacing

\bibitem{BOTTA20123531}
\BIBentryALTinterwordspacing
A.~Botta, A.~Dainotti, and A.~Pescapé, ``A tool for the generation of realistic network workload for emerging networking scenarios,'' \emph{Computer Networks}, vol.~56, no.~15, pp. 3531--3547, 2012. [Online]. Available: \url{https://www.sciencedirect.com/science/article/pii/S1389128612000928}
\BIBentrySTDinterwordspacing

\end{thebibliography}


\begin{thebibliography}{1}
\bibliographystyle{IEEEtran}

\bibitem{ref1}
{\it{Mathematics Into Type}}. American Mathematical Society. [Online]. Available: https://www.ams.org/arc/styleguide/mit-2.pdf

\bibitem{ref2}
T. W. Chaundy, P. R. Barrett and C. Batey, {\it{The Printing of Mathematics}}. London, U.K., Oxford Univ. Press, 1954.

\bibitem{ref3}
F. Mittelbach and M. Goossens, {\it{The \LaTeX Companion}}, 2nd ed. Boston, MA, USA: Pearson, 2004.

\bibitem{ref4}
G. Gr\"atzer, {\it{More Math Into LaTeX}}, New York, NY, USA: Springer, 2007.

\bibitem{ref5}M. Letourneau and J. W. Sharp, {\it{AMS-StyleGuide-online.pdf,}} American Mathematical Society, Providence, RI, USA, [Online]. Available: http://www.ams.org/arc/styleguide/index.html

\bibitem{ref6}
H. Sira-Ramirez, ``On the sliding mode control of nonlinear systems,'' \textit{Syst. Control Lett.}, vol. 19, pp. 303--312, 1992.

\bibitem{ref7}
A. Levant, ``Exact differentiation of signals with unbounded higher derivatives,''  in \textit{Proc. 45th IEEE Conf. Decis.
Control}, San Diego, CA, USA, 2006, pp. 5585--5590. DOI: 10.1109/CDC.2006.377165.

\bibitem{ref8}
M. Fliess, C. Join, and H. Sira-Ramirez, ``Non-linear estimation is easy,'' \textit{Int. J. Model., Ident. Control}, vol. 4, no. 1, pp. 12--27, 2008.

\bibitem{ref9}
R. Ortega, A. Astolfi, G. Bastin, and H. Rodriguez, ``Stabilization of food-chain systems using a port-controlled Hamiltonian description,'' in \textit{Proc. Amer. Control Conf.}, Chicago, IL, USA,
2000, pp. 2245--2249.

\end{thebibliography}

\appendices
\section{Proof of property 3}
\textbf{Property 3:}\par
\textit{The bisection bandwidth of a MODRIC container is \((2x + 4q) \cdot C\), where \(q\) is the number of EPSes per EPSC and \(C\) is the capacity of S1 links.}

\textit{The bisection bandwidth of a MODRIC network of dimension \(m \times n\) (where \(1 < m \leq n\)) is \(\mathcal{O}(n \cdot m^2 \cdot C)\).} \par

\textbf{Proof:}\par
While proving this property, we show the canonical version of the network where all the \(q\) layers of S1 boundary links between any pair of EPSes are shown as one. Hence, each such link has a capacity of \(q \cdot C\). However, EPS to AS links are taken as having a capacity of only \(C\), since only one S1 cable is used to connect a pair of EPS and AS.

\subsection{Bisection width of a MODRIC Container:}
Each of the two equal partitions in a MODRIC container will contain two of the EPSes and \(x/2\) ASes. \par

\begin{figure}[!t]
        \begin{center}
            \includegraphics[width=3.2in]{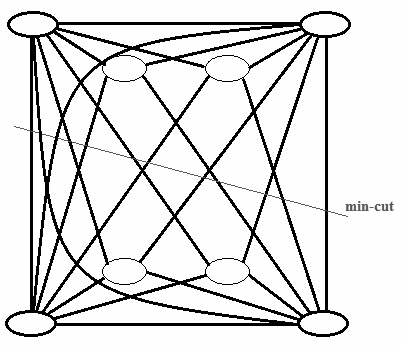}
            \caption{Min-cut bisecting a MODRIC container with 4 ASes}
            \label{fig_15}
        \end{center}
\end{figure}

The min-cut will cut the following links (Figure \ref{fig_15}):
\begin{itemize}
    \item \((2 \times \frac{x}{2})\) links from the \(\frac{x}{2}\) ASes in the upper partition to the 2 EPSes in the lower partition.
    \item \((2 \times \frac{x}{2})\) links from the \(\frac{x}{2}\) ASes in the lower partition to the 2 EPSes in the upper partition.
    \item \(4q\) boundary links between the EPSes in the two partitions.
\end{itemize}

Thus, there will be \(2x + 4q\) links, each having a bandwidth of \(C\). Hence, the MODRIC container bisection bandwidth is \((2x + 4q) \cdot C\).

\begin{figure}[!t]
        \begin{center}
            \includegraphics[width=3.2in]{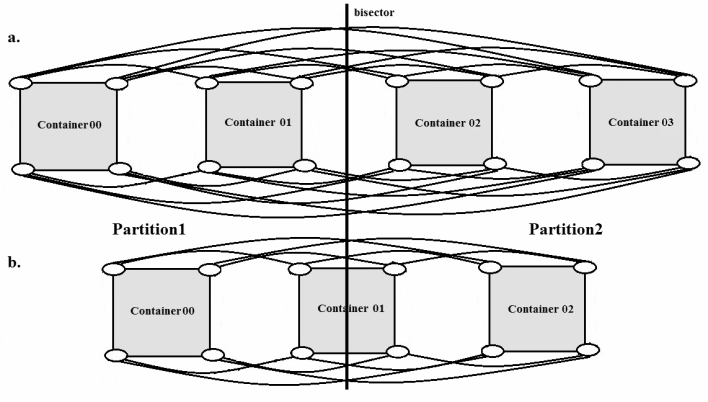}
            \caption{A vertical cut bisecting (a) 1x4 MODRIC (even number of containers in a row) and (b) 1x3 MODRIC (odd number of containers in a row)}
            \label{fig_16}
        \end{center}
\end{figure}

\begin{figure}[!t]
        \begin{center}
            \includegraphics[width=3.2in]{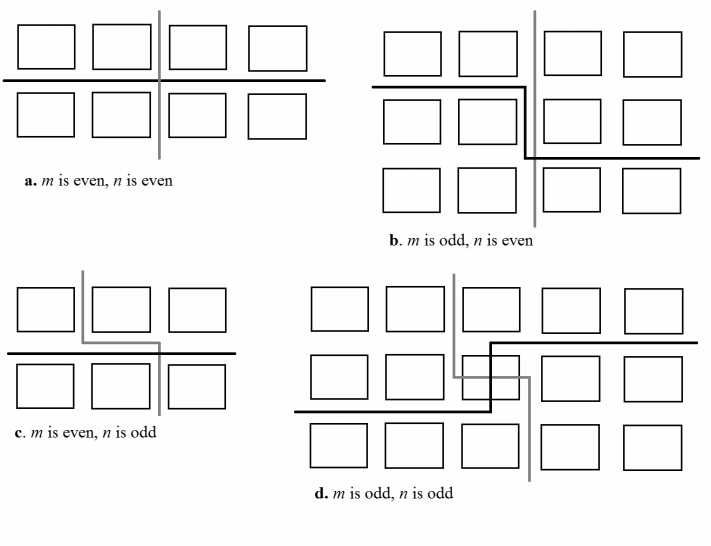}
            \caption{Horizontal and vertical bisectors in the cases where (a) \(m\) and \(n\) both are even, (b) \(m\) is odd and \(n\) is even, (c) \(m\) is even and \(n\) is odd, and (d) \(m\) and \(n\) both are odd.}
            \label{fig_17}
        \end{center}
\end{figure}

\subsection{Bisection of a Column/ Row:}
\textbf{Case1: m, n even:}\par
Consider the bi-partitioning of a row. As shown in Fig. 16a, the partition will have \(n/2\) containers in each half. The partitioning will avoid cutting any of the intra-container links. The inter-container links that will be cut will be the 4 links from each of the \(n/2\) containers in partition 1 to each of the \(n/2\) containers in partition 2, with a total of \(4 \cdot \frac{n^2}{4} = n^2\) links. Each link is of capacity \(C\). The bisection bandwidth of the row will thus be \(n^2 \cdot C\).

Due to symmetry in construction, the bisection bandwidth of a column will be \(m^2 \cdot C\).

\textbf{Case2: m, n odd:}\par
In this case, one of the containers will also have to be partitioned as shown in Fig. 16b. The number of links cut here will be
\[
4 \cdot \left(\left\lfloor \frac{n}{2} \right\rfloor\right)^2 + 4 \cdot \left\lfloor \frac{n}{2} \right\rfloor + (2x + 4q).
\]
The first term is due to the inter-partition inter-container links between the undivided containers, the second term is due to the inter-partition links from the two halves of the partitioned container, and the third term is due to the inter-partition intra-container links.  
Thus, the bisection bandwidth of a row with \(n\) odd is 
\[
\left(4 \cdot \left(\left\lfloor \frac{n}{2} \right\rfloor\right)^2 + 4 \cdot \left\lfloor \frac{n}{2} \right\rfloor + (2x + 4q)\right) \cdot C.
\]
Similarly, the bisection bandwidth of a column with \(m\) odd is 
\[
\left(4 \cdot \left\lfloor \frac{m}{2} \right\rfloor^2 + 4 \cdot \left\lfloor \frac{m}{2} \right\rfloor + (2x + 4q)\right) \cdot C.
\]

\subsection{Bisection of a (m x n) MODRIC Network:}

A min-cut for equal bi-partition of the MODRIC network shall be one that avoids cutting through the containers as that will lead to cutting the intra-container links over and above the inter-container links. In the cases of m even or n even it is possible to have such partitions with a horizontal or a vertical cut respectively.

\textbf{Case1: m even, n even:}\par
A horizontal bi-partitioning can be done with \(m/2\) rows in each of the partitions as shown in Figure \ref{fig_17}a. The partition will cut all the \(n\) columns of the network. Hence, the cut bandwidth will be \(n \cdot m^2 \cdot C\).

A vertical bi-partitioning can similarly be done for a cut bandwidth of \(m \cdot n^2 \cdot C\).

As \(m \leq n\), the min-cut will be horizontal, leading to a bisection width of \(n \cdot m^2 \cdot C\) for the network.

\textbf{Case2: m odd, n even:}\par
In this case, the vertical cut will have a bandwidth of 
\[
m \cdot n^2 \cdot C = (n \cdot C) \cdot m \cdot n \tag{1}
\]
A horizontal bi-partitioning avoiding partitioning of any of the containers is also possible with the middle row bi-partitioned with a vertical cut (Figure \ref{fig_17}b). The bandwidth of the cut in this case will be 
\[
n \cdot m^2 \cdot C + n^2 \cdot C = n \cdot C \cdot (m^2 + n) \tag{2}
\]
Considering the dominant term in (2), the bisection bandwidth will be \(n \cdot m^2 \cdot C\).

\textbf{Case3: m even, n odd:}\par
In this case (Figure \ref{fig_17}c), the min-cut will be a horizontal one, as in case 1, with a bisection bandwidth of \(n \cdot m^2 \cdot C\).

\textbf{Case4: m odd, n odd:}\par
In this case, one of the containers of the middle row will have to be bisected as shown in Figure \ref{fig_17}d into a horizontal cut of the \(n\) columns. The bisection bandwidth will be  
\[
n \cdot m^2 \cdot C + \left(4 \cdot \left\lfloor \frac{n}{2} \right\rfloor^2 + 4 \cdot \left\lfloor \frac{n}{2} \right\rfloor + (2x + 4q)\right) \cdot C.
\]
Considering the dominant terms in all the four cases, the bisection bandwidth of a \((m \times n)\) MODRIC network is \(\mathcal{O}(n \cdot m^2 \cdot C)\).

\end{document}